\documentclass[aps, pra, reprint, superscriptaddress,longbibliography]{revtex4-2}

\usepackage{lineno,hyperref}
\usepackage{eurosym}
\usepackage{amsmath,amssymb,graphicx}
\usepackage{color}
\usepackage{float}
\usepackage[normalem]{ulem}

\modulolinenumbers[5]
\newcommand{\bitem}{\begin{itemize}}
\newcommand{\fitem}{\end{itemize}}
\newcommand{\beq}{\begin{equation}}
\newcommand{\eeq}{\end{equation}}
\newcommand{\beqa}{\begin{eqnarray}}
\newcommand{\eeqa}{\end{eqnarray}}

\begin{document}

\title{\textbf{Quantum Otto Heat Engine based on the Dicke-Stark Model under Infinite-Time and Finite-Time Thermodynamic Frameworks} 
}%
\author{Weilin Wang}
\affiliation{Department of Mathematics and Physics, North China Electric Power University, Huadian Road, Baoding 071000, China}
\author{Xiyuan Zhang}
\affiliation{Department of Mathematics and Physics, North China Electric Power University, Huadian Road, Baoding 071000, China}
\author{Weiran  Zhao}
\affiliation{Department of Mathematics and Physics, North China Electric Power University, Huadian Road, Baoding 071000, China}
\author{Mingshu Zhao}
\affiliation{Department of Mathematics and Physics, North China Electric Power University, Huadian Road, Baoding 071000, China}
\author{Jinying Ma}
\affiliation{Department of Mathematics and Physics, North China Electric Power University, Huadian Road, Baoding 071000, China}
\author{Zhanyuan Yan}
\email{Contact author:yanzhanyuan@ncepu.edu.cn}
\affiliation{Department of Mathematics and Physics, North China Electric Power University, Huadian Road, Baoding 071000, China}
\affiliation{Hebei Key Laboratory of Physics and Energy Technology, North China Electric Power University, Baoding 071003, China}

\date{\today}

\begin{abstract}
We propose a quantum Otto heat engine that employs a finite-size Dicke--Stark model as the working substance. In the extended coherent state space, the complete energy spectrum and eigenstates of this model are obtained through numerical calculations. Within the infinite-time and finite-time thermodynamics frameworks, we investigate the effects of the Stark field strength, coupling strength, adiabatic stroke time, isochoric stroke time, and number of atoms in the DS model on the heat engine's output work, efficiency, and power.
The results show that the maximum values of the output work and efficiency appear near the coupling strength corresponding to the superradiant phase transition point. Regulating the Stark field strength can tune the energy level structure of the system and the superradiant phase transition, effectively reducing entropy generation and quantum friction during nonequilibrium evolution of the system's states and thereby significantly increasing the engine's output work, efficiency, and power. Asymmetric heat engines, where the two isochoric strokes have different Stark field strengths and stroke times, are more conducive to optimizing the heat engine's performance. Additionally, in the DS model, an increase in the number of atoms is also beneficial for increasing the heat engine's output work and efficiency. The results of this paper facilitate the design of high-performance quantum heat engines.

\end{abstract}
\maketitle
\section{\label{sec:level1}INTRODUCTION}
Quantum heat engines have recently attracted significant attention due to their potential as microscale energy conversion devices and their promising applications in nanotechnology \cite{geusic1967quantum,kosloff2013quantum}. Unlike conventional heat engines, quantum heat engines employ quantum systems as their working substance, such as cavity quantum electrodynamics (QED) systems that describe the interaction between qubits and quantum light fields \cite{dodonov2018quantum,kerremans2022probabilistically,wang2009thermal,buffoni2019quantum}. Consequently, these engines can leverage the quantum properties of the system, including entanglement and coherence, to achieve performance that reaches or even exceeds traditional thermodynamic limits~\cite{rossnagel2014nanoscale,klaers2017squeezed}. Experimentally, quantum heat engines have been realized on various platforms, including trapped-ion systems \cite{maslennikov2019quantum}, optomechanical platforms \cite{zhang2014quantum,zhang2014theory}, ultracold atomic ensembles \cite{brantut2013thermoelectric} and superconducting circuits \cite{koski2015chip,pekola2016maxwell}.

The thermodynamic cycle is central to the design of quantum heat engines, with typical cycles including the Carnot cycle, Otto cycle, and Stirling cycle \cite{kosloff2017quantum,deffner2019quantum,thomas2017implications,thomas2017implications,raja2021finite}. Among these, the quantum Otto cycle is a primary research focus due to its simplicity and ease of quantization. This cycle comprises two isochoric processes and two adiabatic processes:
During isochoric processes, the working substance exchanges heat with a heat reservoir, whereas in adiabatic processes, the energy levels of the working substance are adjusted by an external field. Efficiency and output work serve as key metrics for evaluating the performance of a heat engine.
Previous studies often assumed quasistatic operation with an infinite cycle time, which achieves an efficiency close to the Carnot limit but results in zero power output \cite{thomas2014friction, alecce2015quantum, plastina2014irreversible}. However, practical applications require balancing efficiency and power within a finite time, and this challenge has driven the development of finite-time quantum thermodynamics, highlighting the crucial impact of quantum friction and nonadiabatic transitions \cite{abah2019shortcut, shiraishi2016universal, keller2020feshbach} on engine performance.

Enhancing the performance of a quantum heat engine involves two key strategies:

First, the energy output is maximized through the selection of the working substance. The working substances currently under extensive investigation include , oscillator systems \cite{anka2021measurement}, spin systems \cite{myers2021quantum}, quantum gases \cite{ccakmak2019spin, peterson2019experimental}, and many-body quantum systems \cite{fogarty2020many}. As a representative of many-body systems, the Dicke model describes the collective coupling of $N$ two-level atoms to a single-mode light field \cite{dicke1954coherence}. A crucial characteristic of the Dicke model is the superradiant phase transition (SPT): When the coupling strength exceeds a critical threshold, the ground state of the system undergoes an abrupt transition from a normal phase characterized by a low photon number to a superradiant phase, marked by macroscopic photon occupation and collective atomic polarization \cite{hioe1973phase, wang1973phase}. The associated critical fluctuations and many-body entanglements can significantly increase the energy transport efficiency \cite{kim2025observation, kumar2025enhancing}.

Second, the energy losses are minimized via nonequilibrium control. Recent developments in finite-time quantum thermodynamics indicate that nonequilibrium dissipation can be effectively prevented by optimizing external field control protocols, such as shortcuts to adiabaticity (STAs) \cite{abah2019shortcut, guery2019shortcuts, del2018friction, torrontegui2013shortcuts,abiuso2020optimal},
and the energy transfer efficiency can be increased through counterdiabatic driving fields that effectively prevent nonequilibrium dissipation \cite{campo2014more, del2013shortcuts}.
Additionally, the performance of many-body heat engines is sensitive to the superradiant phase transition. Controlling the superradiant phase transition point with an external field can effectively reduce quantum friction and increase the efficiency of a heat engine \cite{fogarty2020many, piccitto2022ising,ma2017quantum,chen2019interaction,b2020universal,revathy2024improving}.

Based on the above two strategies, the Dicke--Stark (DS) model has emerged as a promising candidate for working substances. By applying an external electric field, the energy difference of the two-level atom and the superradiant phase transition point in this model can be modified. Rational control of the external electric field, nonequilibrium quantum dissipation and quantum friction can be effectively reduced, thereby improving the performance of the heat engine.

The solution from the DS model also presents a significant challenge. When the atom--field coupling strength reaches ultrastrong or deep--strong coupling regimes, the rotating wave approximation (RWA) is no longer valid \cite{chen2012spectrum,fink2008climbing}. In such cases, only partial results such as ground state and phase transitions can be obtained at the thermodynamic limit (where the number of atoms $N \to \infty $) by exploiting the collective symmetry of the Dicke system \cite{chen2008numerically,emary2003quantum}. However, the calculation of the overall energy spectra and eigenstates of the infinite-size DS model remains an unsolved issue. Consequently, the finite-size DS model has become a viable working substance for research, particularly within the coherent-state basis where computational demands are substantially reduced, enabling efficient calculation of the system's energy spectrum and corresponding wavefunctions. Furthermore, in the strong coupling regimes, only the dressed-state solution from the model can describe the quantum state of the system. Therefore, formulating and solving the dressed master equation constitutes the foundation for studying quantum heat engines within the framework of finite-time thermodynamics \cite{le2016fate,ye2021quantum,settineri2018dissipation}.

In this work, we employ the Dicke--Stark model as the working substance to investigate a four-stroke quantum Otto cycle heat engine. The remainder of this paper is organized as follows: In Section II, the dressed master equation of the quantum DS model and its numerical solution methods on a coherent-state basis are presented. In Section III, the working cycle process of the four-stroke quantum heat engine is introduced. The optimization of heat engine performance under the infinite-time and finite-time thermodynamic frameworks is investigated in detail in Sections IV and V, respectively.
The main findings are summarized in Section VI.

\section{\label{sec:level2}Models and Methods}
\subsection{\label{sec:level2-1}Dicke--Stark Model}
The Dicke--Stark model describes a quantum system composed of $N$ two-level qubits interacting with a single-mode bosonic field. The Hamiltonian of the system can be expressed as ($ \hbar = 1$\ and $ k _ { B } = 1$)
\cite{garraway2011dicke,gopalakrishnan2011frustration,
bastidas2012nonequilibrium,abdel2017quantum,mu2020dicke}
:
\begin{equation}\label{H_D}
{\hat H_{DS}} = \omega {\hat a^ \dagger  }\hat a + \Delta {\hat J_z} + \frac{{2\lambda }}{{\sqrt N }}({\hat a^ \dagger  } + \hat a){\hat J_x} + \frac{U}{N}{\hat a^ \dagger  }\hat a{\hat J_z},
\end{equation}
where ${\hat J_x} $ and $\hat { J } _ { z }$ represent the collective
spin operators, composed of $\hat { J } _ { \alpha } = \sum _ { i=1 } ^ { N } \frac{1}{2}\hat { \sigma } _ { \alpha } ^ { i }$, with $ \hat { \sigma } _ { \alpha } ( \alpha = x, y, z)$ as the Pauli operators. ${\hat a^ \dagger  }$ and ${\hat a }$ denote the creation and annihilation operators of the bosonic field. $\Delta $ and $\omega $ represent the frequency of the qubits and the single bosonic mode, and $\lambda$ is the qubit--boson coupling strength. $\ U$ represents the intensity of the nonlinear Stark interaction. In the following sections, we set the bosonic frequency $\omega$ as the energy unit for simplicity.

Owing to the unavailability of complete eigenenergy and eigenstate information for the DS model in the thermodynamic limit ($N \to \infty $), we select the finite-size DS model as the working substance for our heat engine. Numerically calculating the eigenenergies and eigenstates of the finite-size DS model in the Fock state space of the bosonic field requires a large photon number cutoff to ensure computational accuracy; this leads to the challenge of solving a Hamiltonian matrix with enormous dimensions, which poses a great challenge to computer memory and computation time.
\subsection{\label{sec:level2-2}Extended Coherent Bosonic State Approach}
The extended coherent bosonic state approach is proposed for accurately calculating the eigenenergies and eigenstates of the finite-size Dicke model using a small photon number cutoff
\cite{chen2008numerically}.
This method is also applicable to the finite-size DS model \cite{wang2025}, making it possible to study heat engines with the DS model as the working substance. Before including the extended coherent bosonic state method, we rotate the collective spin operators with $\pi /2$ along $\hat{J}_{y}$ by ${\hat H} = \exp(i\pi {\hat J_y}/2){\hat H_D}\exp( - i\pi {\hat J_y}/2)$, resulting in
\begin{equation}\label{H}
\hat H = \omega {\hat a^\dagger }\hat a - (\frac{\Delta }{2} + \frac{U}{{2N}}{\hat a^\dagger }\hat a)({\hat J_ + } + {\hat J_ - }) + \frac{{2\lambda }}{{\sqrt N }}({\hat a^\dagger } + \hat a){\hat J_z}.
\end{equation}
in which ${\hat { J } _ { \pm } = \hat {J} _ {x} \pm  \hat {J} _ {y} }$ are the raising and lowering operators of the atomic states $\{ \left| {j,m} \right\rangle,m =  - j, - j + 1, \ldots,j - 1,j \}$ with $j = N/2$. Then, the states in the Hilbert space of the entire system can be represented by the direct product of the bosonic field states and the atomic states, $\left\{ {|{\varphi^n}{ \rangle _b} \otimes |j,m\rangle } \right\}$.

Considering a displacement operator $\hat{D}(g_m)=\exp(g_m\hat{a}-g_m\hat{a}^{\dagger})$, with ${g_m} = 2\lambda m/\omega \sqrt N$, the displaced operators ${\hat{A}_m} =\hat{D}^{\dagger} \hat{a}\hat{D}=\hat{a} + {g_m}$, ${\hat{A}^{\dagger}_m} =\hat{D}^{\dagger} \hat{a}^{\dagger}\hat{D} =\hat{a}^{\dagger} + {g_m}$ serve as the creation and annihilation operators of the extended coherent state space, which is defined as follows:
$$|k{\rangle _{A_m}={\frac{1}{{\sqrt {k!} }}} (\hat A_m^ \dagger  )^k}|0{\rangle _{A_m}},$$  where $|0{\rangle} _{A_m}=\hat{D}^ \dagger|0\rangle_a$ is the vacuum state in the extended coherent state space.

Then, the photonic state can be expanded in the extended coherent state space as
\begin{equation*}
|{\varphi^n_m}{\rangle _b} = \sum\limits_{k = 0}^{N_{tr}} {\frac{1}{{\sqrt {k!} }}} {C^n_{m,k}}{(\hat A_m^ \dagger  )^k}|0{\rangle _{A_m}}
\end{equation*}
Obviously, the expansion of the photonic state in the extended coherent state space includes all Fock states; thus, a relatively small truncation number $N_{tr}$ can yield accurate calculation results. Substituting the state of the entire system into the Schrodinger equation yields the following:
\begin{widetext}
\begin{equation}
\begin{array}{l}
\sum\limits_{m,k} {\omega C_{m,k}^n} (k - g_m^2)|j,m\rangle|k{\rangle_{{A_m}}}\\
 - \sum\limits_{m,k} {C_{m,k}^n} \left[ {\frac{\Delta }{2} + \frac{U}{{2N}}(k + g_m^2)} \right](j_m^ + |j,m + 1\rangle|k{\rangle _{{A_m}}}
 + j_m^ - |j,m - 1\rangle |k{\rangle _{{A_m}}})\\
 + \frac{U}{{2N}}\sum\limits_{m,k} {C_{m,k}^n} {g_m}(\sqrt {k + 1} j_m^ + |j,m + 1\rangle|k + 1{\rangle _{{A_m}}}+ \sqrt k j_m^ + |j,m + 1\rangle |k - 1{\rangle _{{A_m}}})\\
 + \frac{U}{{2N}}\sum\limits_{m,k} {C_{m,k}^n} {g_m}(\sqrt {k + 1} j_m^ - |j,m - 1\rangle |k + 1{\rangle _{{A_m}}}+ \sqrt k j_m^ - |j,m - 1\rangle |k - 1{\rangle _{{A_m}}})\\
 = {E_n}\sum\limits_{m,k} {C_{m,k}^n} |j,m|k{\rangle _{{A_m}}}.
 \end{array}
\end{equation}
\end{widetext}
We then multiply $\left\{ {\left\langle l \right|\left\langle {n,j} \right|} \right\}$ by $n =  - j, - j + 1, \ldots ,j$. We obtain the equation satisfied by the expansion coefficients $C^n_{m,k}$
\begin{equation}
\begin{array}{l}
\omega C_{n,l}^n(k - g_n^2)\\
 - \left[ {\frac{\Delta }{2} + \frac{U}{{2N}}(k + g_{_{n - 1}}^2)} \right]C_{_{n - 1},l}^nj_{_{n - 1}}^ + \sum\limits_k {{}_{{A_n}}\left\langle l | k \right\rangle }_{{A_{n - 1}}} \\
 + \left[ {\frac{\Delta }{2} + \frac{U}{{2N}}(k + g_{_{n + 1}}^2)} \right]C_{_{n + 1},l}^nj_{_{n + 1}}^ - \sum\limits_k {{}_{{A_n}}\left\langle l|k \right\rangle }_{{A_{n + 1}}} \\
 + \frac{U}{{2N}}C_{n - 1,k}^n{g_{n - 1}}\sqrt {k + 1} j_{n - 1}^ + \sum\limits_k {{}_{{A_n}}\left\langle l|{k + 1} \right\rangle }_{{A_{n - 1}}} \\
 + \frac{U}{{2N}}C_{n - 1,k}^n{g_{n - 1}}\sqrt k j_{n - 1}^ + \sum\limits_k {{}_{{A_n}}\left\langle l | {k - 1} \right\rangle }_{{A_{n - 1}}} \\
 + \frac{U}{{2N}}C_{n + 1,k}^n{g_{n + 1}}\sqrt {k + 1} j_{n + 1}^ - \sum\limits_k {{}_{{A_n}}\left\langle l | {k + 1} \right\rangle }_{{A_{n + 1}}} \\
 + \frac{U}{{2N}}C_{n + 1,k}^n{g_{n + 1}}\sqrt k j_{n + 1}^ - \sum\limits_k {{}_{{A_n}}\left\langle l | {k - 1} \right\rangle }_{{A_{n + 1}}} \\
 = {E_n}C_{n,l}^n,
\end{array}
\end{equation}
in which the inner products are
\begin{equation}
    _{A_n}{\langle l|k\rangle _{{A_{n - 1}}}} = {( - 1)^l}{D_{l,k}},_{A_n}{\langle l|k\rangle _{{A_{n + 1}}}} = {( - 1)^k}{D_{l,k}}, \notag
\end{equation}
\begin{equation}
{D_{l,k}} = {e^{ - {G^2}/2}}\sum\limits_{r = 0}^{\min (l,k)} {\frac{{{{( - 1)}^{ - r}}\sqrt {l!k!} {G^{l + k - 2r}}}}{{(l - r)!(k - r)!r!}}} , G = \frac{{2\lambda }}{{\omega \sqrt N }}. \notag
\end{equation}

Notably, the theoretical range for the Stark field strengthh parameter \(U\) is \([-\infty, 2\omega]\) \cite{wang2025}. However, considering experimental feasibility and the need for many photons to obtain accurate energy levels as \(U\) increases, in the subsequent research, we limited the range of \(U\) to \(|U| < \omega\) and adopted a maximum truncation number \(N_{tr} = 60\). This parameter setting is sufficient to ensure that the calculated excitation state energies converge to the target accuracy within the \(|U| < \omega\) region, with a relative error less than \(10^{-4}\).

\subsection{\label{sec:level2-3}Quantum Dressed Master Equation}
According to the method outlined in \cite{mccauley2020accurate,xu2024universal,xu2024exploring},
we can derive the thermodynamically effective master equation for the entire Hamiltonian spectrum, which holds under weak damping conditions:
A high reservoir cutoff frequency and a flat spectral density ensure that the system is positive and has Markovian properties. Under these conditions, the system dynamics are governed by the dressed-state master equation
\cite{gorini1976completely,lindblad1976generators,weiss2012quantum,breuer2002theory,le2016fate,ye2021quantum,settineri2018dissipation}
\begin{eqnarray}~\label{eq:master_equation}
	\frac{d}{dt}\hat{\rho}&=&-i[\hat{H},\hat{\rho}]+\sum_{ \substack{u=a,\sigma^{-}\\k<j } }
	\{\Gamma^{jk}_un_u(\Delta_{jk}){D}[|\phi_j{\rangle}{\langle}\phi_k|,\hat{\rho}]\nonumber\\
	&&+\Gamma^{jk}_u[1+n_u(\Delta_{jk})]{D}[|\phi_k{\rangle}{\langle}\phi_j|,\hat{\rho}]\},\label{eq:dressed-me}
\end{eqnarray}
where $D\left[ {\hat O,\hat \rho } \right] = \hat O{\hat O^ \dagger  } - \frac{1}{2}\left\{ {{{\hat O}^ \dagger  }\hat O,\hat \rho } \right\}$. The dissipation rates $\Gamma _u^{jk} = {\gamma _u}({\Delta _{jk}})|S_u^{jk}{|^2}$ depend on the spectral function ${\gamma _u}({\Delta _{jk}})$ and the transition coefficients $S_a^{jk} = \langle {\phi _j}|({\hat a^\dagger } + \hat a)|{\phi _k}\rangle $ and $S_{\sigma  - }^{jk} = \langle {\phi _j}|({\hat \sigma _ + }+ {\hat \sigma _ - })|{\phi _k}\rangle $. For the ohmic case, ${\gamma _u}({\Delta _{jk}}) = \pi \alpha {\Delta _{jk}}\exp( - |{\Delta _{jk}}|/{\omega _c})$, where $\alpha $ represents the coupling strength and where ${\omega _{{c}}}$ represents the cutoff frequency; throughout all the numerical simulations performed, we consider $\alpha  = 0.001\omega $ and ${\omega _c} = 10\omega $. The Bose--Einstein distribution ${n_u}({\Delta _{jk}},{T_u}) = 1/\left[ {\exp({\Delta _{jk}}/{T_u}) - 1} \right]$ accounts for thermal effects. The steady-state solution to Eq.~\eqref{eq:master_equation} yields the density matrix of the canonical ensemble, as confirmed by straightforward numerical simulations:
\begin{equation}\label{eq:rho}
    {\hat \rho _{ss}} = \sum\limits_n {\frac{{{e^{ - {E_n}/T}}}}{Z}} |{\phi _n}\rangle \langle {\phi _n}|.
\end{equation}
where $Z = \sum\limits_n {{e^{ - {E_n}/T}}} $ represents the partition function and  $P_n^{ss}=\frac{{{e^{ - {E_n}/T}}}}{Z}$ is the population.

\section{\label{sec:level3}
Quantum Otto cycle
}
\begin{figure}[htbp]
\includegraphics[width=0.48\textwidth]{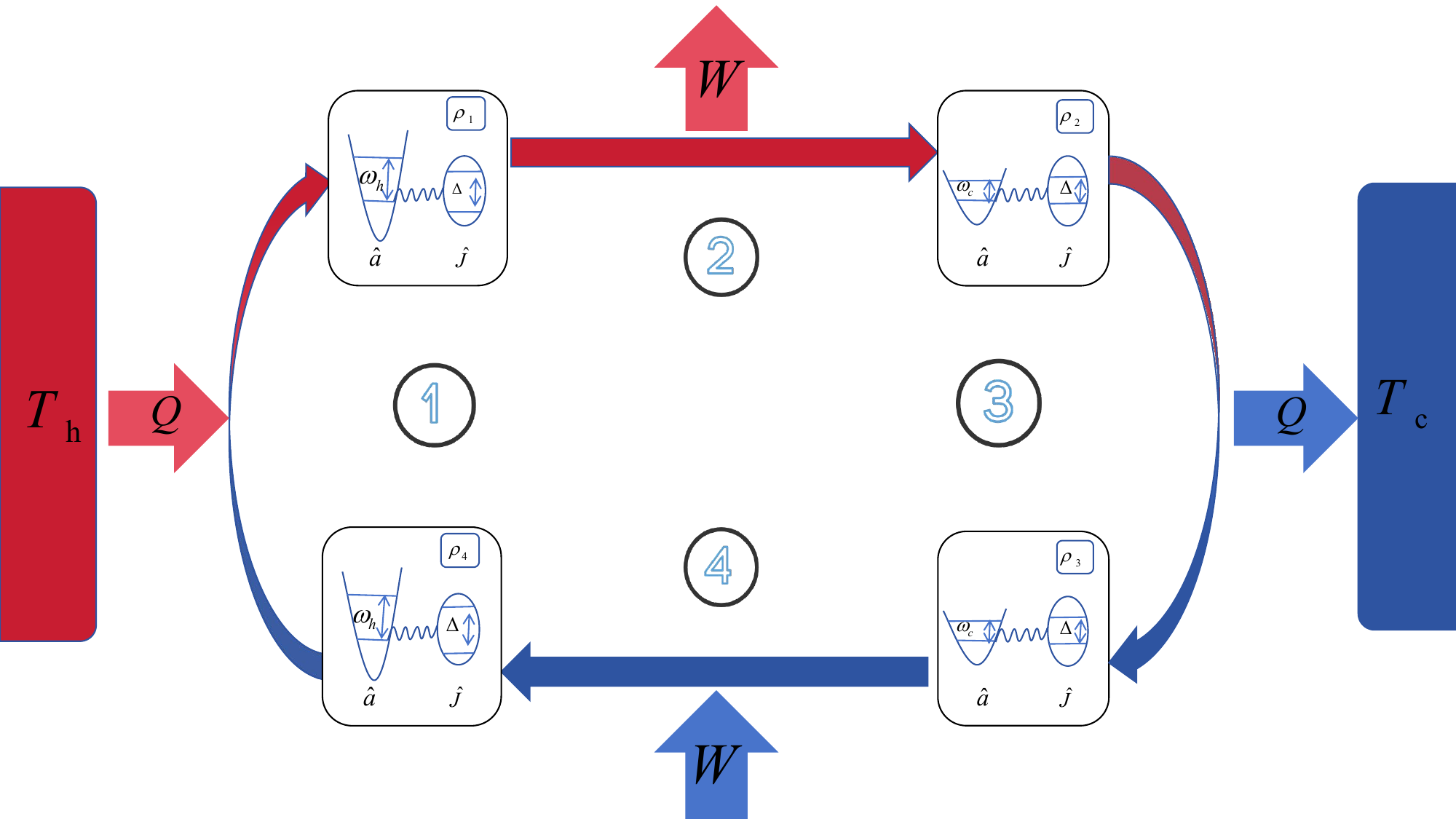}
\caption{Schematic diagram of a four-stroke Otto cycle quantum heat engine based on the Dicke--Stark model. Processes 1 (3) represent isochoric heating (cooling), where the working substance interacts with a hot (cold) reservoir at temperature $T_h(T_c)$. The frequency remains constant, heat is absorbed (released), and no work is performed.
Process 2 (4) represents adiabatic expansion (compression), where the working substance is isolated from heat reservoirs. The frequency is changed from $\omega_h$($\omega_c$) to $\omega_c$ ($\omega_h$). The system energy shifts from $E_h$($E_c$) to $E_c$($E_h$). No heat is exchanged, and the system performs positive (negative) work on the outside. }~\label{otto_Otto cycle}
\end{figure}
The quantum Otto engine works by cyclically manipulating the parameters of the working substance, including four main stages: isochoric heating, adiabatic expansion, isochoric cooling and adiabatic compression
\cite{abah2019shortcut}.
The thermodynamic cycle that we consider is shown in Fig.~\ref{otto_Otto cycle}.

\textit{\textbf{Quantum isochoric heating process:}}
A working substance characterized by Hamiltonian $H^{h}$ and frequency $\omega_{h}$ contacts a hot reservoir at temperature $(T_h)$ and gradually heats. The state of the system evolves from $\rho(T_c)$ to $\rho(T_h)$ without performing any work, and the heat absorbed during this process is $Q_h$.

\textit{\textbf{Quantum adiabatic expansion process:}}

The system is isolated from heat reservoirs. By changing its frequency from \(\omega_h\) to \(\omega_c\) (\(\omega_h > \omega_c\)), the energy levels change from \(E_n^h\) to \(E_n^c\). During this process, output work $W$ is performed without heat exchange.

\textit{\textbf{Quantum isochoric cooling process:}}
The working substance, with a frequency of ${\omega _c}$, contacts a cold reservoir at temperature ${T_c} < {T_h}$.
The state of the system evolves from $\rho(T_h)$ to $\rho(T_c)$ without performing any work, and the heat absorbed during this process is $Q_c$.

\textit{\textbf{Quantum adiabatic compression process:}} Isolated from the cold reservoir, the energy levels of the system change from $E_n^c$ to $E_n^h$ as the frequency changes from ${\omega _c}$ to ${\omega _h}$. Input work $W$ is performed without heat exchange.

A positive value of $\ Q $ indicates that the working substance absorbs heat from the reservoir, whereas a negative value indicates heat release.
Similarly, positive values of $\ W$ indicate the output work of the engine, and negative values indicate input work on the heat engine. The allowed working regimes under the Clausius inequality and the first law of thermodynamics are as follows:

\textit{\textbf{ Heat engine (E): $Q_{{h}} > 0$, $Q_{{c}} < 0$, $W > 0$}};

\textit{\textbf{ Refrigerator (R): $Q_{{c}} > 0$, $Q_{{h}} < 0$, $W > 0$}};

\textit{\textbf{ Heater (H): $Q_{{c}} < 0$, $Q_{{h}} < 0$, $W < 0$}};

\textit{\textbf{ Accelerator (A): $Q_{{c}} < 0$, $Q_{{h}} > 0$, $W < 0$}}.

Among these four operation modes, the heat engine (E) and the refrigerator (R) are of the most practical. In this work, we focus on the heat engine, which operates by absorbing heat from a hot reservoir, converting part of this energy to work, and releasing the remaining heat to a cold reservoir.

\section{\label{sec:level4}Infinite-time Cycle}
\subsection{\label{sec:level4-1}Work and Heat Calculations}
The first law of thermodynamics for quantum systems with discrete energy levels is expressed as follows\cite{kieu2004second}:
\begin{equation}
   dU = \delta Q + \delta W = \sum\limits_n {({E_n}dP_n^{ss} + P_n^{ss}d{E_n})} . 
\end{equation}
where ${E_n}$ represents the energy levels and where $P_n^{ss}$ represents the probabilities of steady-state occupation. The heat ${Q_h}$ and ${Q_c}$ exchange with the hot reservoir and the cold reservoir, respectively, and the net work $W$ is expressed as follows:
\begin{equation}
   {Q_h} = \sum\limits_n {E_n^h} \left[ {P_n^{ss}({T_h}) - P_n^{ss}({T_c})} \right],
\end{equation}
\begin{equation}
   {Q_c} = \sum\limits_n {E_n^c} \left[ {P_n^{ss}({T_c}) - P_n^{ss}({T_h})} \right],
\end{equation}
\begin{equation}\label{eq10}
    W = {Q_h} +  {Q_c} = \sum\limits_n {(E_n^h - E_n^c)} [P_n^{ss}({T_h}) - P_n^{ss}({T_c})].
\end{equation}
The efficiency of the heat engine, $\eta $, is a key figure of merit and is defined as the ratio of the work output to the heat absorbed from the hot reservoir.
\begin{equation}~\label{eq.11}
\eta  = \frac{W}{{{Q_h}}}.
\end{equation}

\subsection{\label{sec:level4-2}Operational Modes in the Infinite-time Cycle}
\begin{figure*}[htbp]
\includegraphics{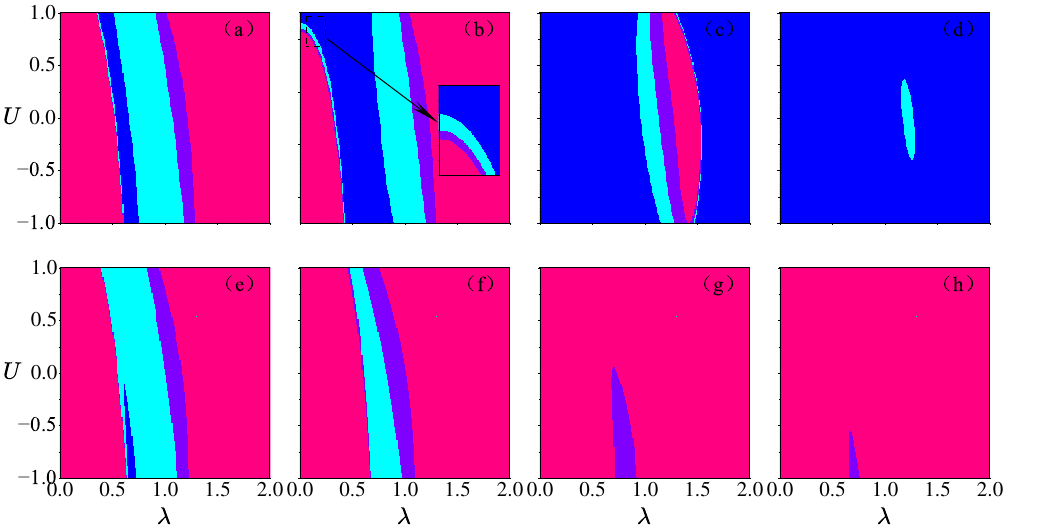}
\caption{Operational state diagram of the heat engine defined by two control parameters: coupling strength (horizontal axis) and Stark field strength (vertical axis). Four distinct operational regions are identified: the heat engine region (pink), the refrigerator region (blue), the accelerator region (light blue), and the eater region (purple). The temperatures are
$(a) \ T_c = 0.1,\ T_h = 0.5$,
$(b) \ T_c = 0.2,\ T_h = 0.5$,
$(c) \ T_c = 0.3,\ T_h = 0.5$,
$(d) \ T_c = 0.4,\ T_h = 0.5$,
$(e) \ T_c = 0.1,\ T_h = 0.6$,
$(f) \ T_c = 0.1,\ T_h = 0.8$,
$(g) \ T_c = 0.1,Th\ = 1.0$,
$(h) T_c  = 0.1,T_h\ = 1.4$, and $N = 8,{\omega _h} = 2\omega ,{\omega _c} = \omega ,\omega = 1$. }\label{otto_operational}
\end{figure*}

We begin by investigating the effects of the reservoir temperature $T$, the coupling strength $\lambda$, and the Stark field strength $U$ on the operational modes of the quantum thermal machine. The phase diagram depicting the operating states of the quantum Otto heat machine is shown in Fig. \ref{otto_operational}. The results show that significant changes occur in the phase diagram of the operational modes when the temperature of the hot reservoir (or cold reservoir) is fixed while the temperature of the cold reservoir (or hot reservoir) is varied. For large temperature differences, the phase diagram is dominated by the heat engine mode [pink regions, Figs.~\ref{otto_operational}(e)-~\ref{otto_operational}(h)]; conversely, for small temperature differences, the refrigeration mode (blue regions) dominates, as shown in Figs.~\ref{otto_operational}(a)-~\ref{otto_operational}(d). Under appropriate coupling strength, tuning the Stark field strength of $U$ enables the heat engine to switch from one operating mode to another. Notably, as shown in the enlarged image in Fig.~\ref{otto_operational}(b), when the coupling strength is approximately $0.1\omega$, precise tuning of $U$ facilitates transitions among four distinct operational modes. These characteristics have substantial potential for developing tunable, multifunctional quantum thermal devices.

\subsection{\label{sec:level4-3} Optimization of Output Work and Efficiency in the Infinite-time Cycle}
\begin{figure*}[htbp]
\includegraphics{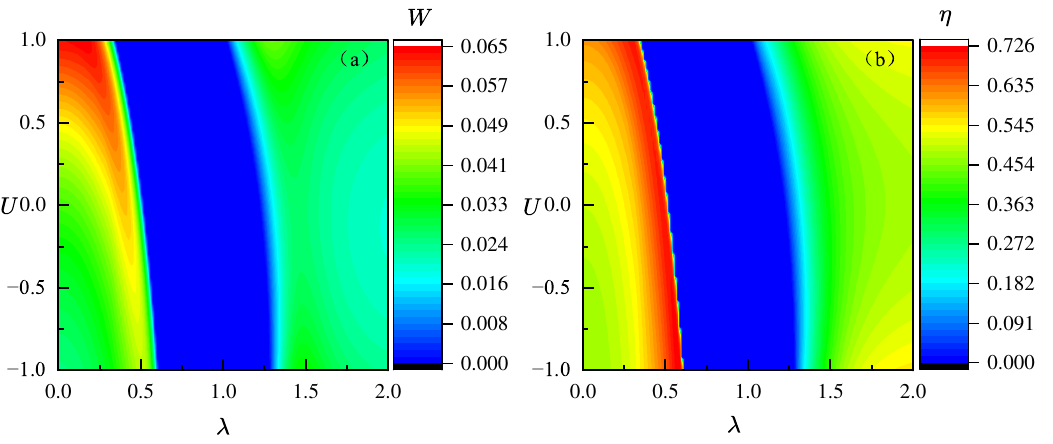}
\caption{Analysis of the two-dimensional heat maps of work output (a) and efficiency (b) for the working substance in the quantum Otto cycle based on the DS model. The diagram quantifies the effects of the Stark field strength ($U$) and coupling strength ($\lambda$) on thermodynamic performance, with color bars indicating the magnitudes of $\ W$ and $\eta$. The additional system parameters are $N = 8$, $\omega _h = 2\omega$, $\omega _c = \omega$, $\omega  = 1$, $T_c = 0.1$, and $ T_h = 0.5$.}~\label{otto_ 2D work and eifficiency}
\end{figure*}

Under quasistatic conditions, the study of the work output and efficiency of a heat engine is crucial for designing heat engines with excellent performance. We use Eq.~\ref{eq10} and Eq.~\eqref{eq.11} to calculate the work output and efficiency under quasistatic conditions. Strategically tuning both the coupling strength $\lambda$ and the Stark field strength $U$ enables the heat engine to achieve optimal work output and efficiency, corresponding to the red shaded regions in Figs.~\ref{otto_ 2D work and eifficiency}(a) and (b), respectively.
\begin{figure*}[htp]
\includegraphics{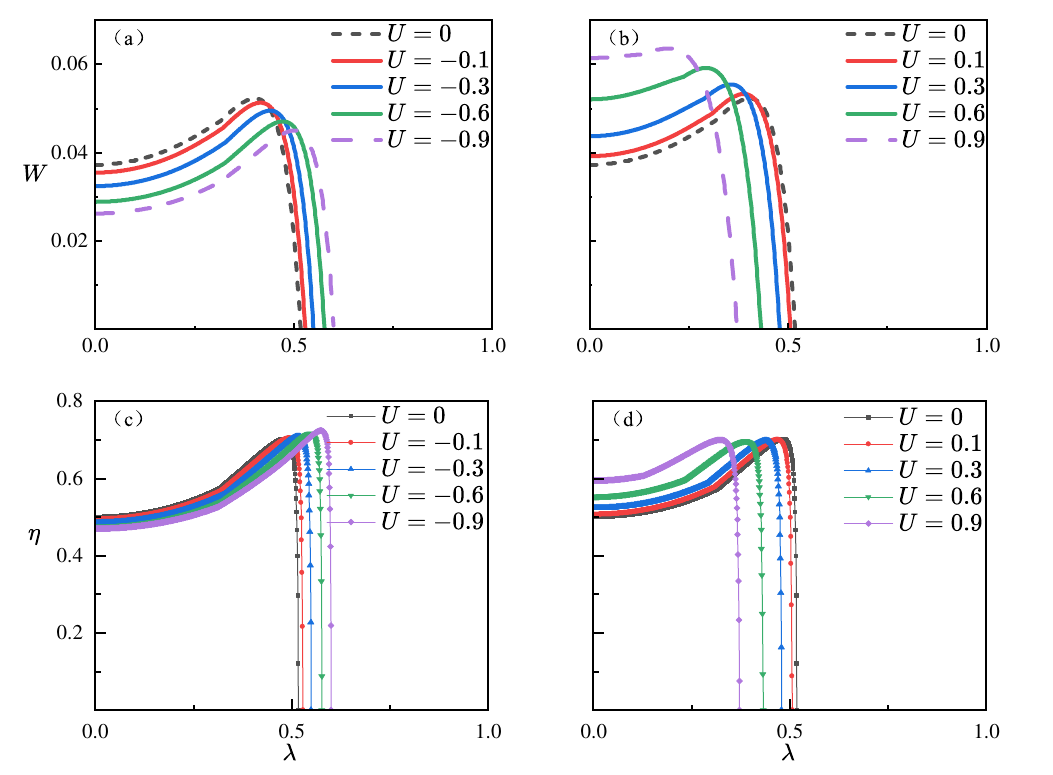}
\caption{Work output (a), (b) and efficiency (c), (d) of a thermal engine as a function of the coupling strength $\lambda$. The horizontal axis denotes $\lambda$, whereas the vertical axis represents the work output ($W$) and efficiency ($\eta$). The curves of different colors represent different Stark field strength $U$ values. The loop parameters include $N = 8$, $\omega _h = 2\omega$, $\omega _c = \omega$, $\omega  = 1$, $T_c = 0.1$, and $ T_h = 0.5$. All the quantities are measured in units of $\omega $. }~\label{otto_work and eifficiency}
\end{figure*}

To elucidate the regulatory effect of $U$ on the output work and efficiency of the heat engine, discrete analyses of $U$ are shown in Fig.~\ref{otto_work and eifficiency}. First, when $U$ is held constant, both the output work $W$ and the efficiency $\eta$ increase gradually with increasing coupling strength $\lambda$, reach a peak and then decrease rapidly.

Next, by comparing the curves of the heat engine's output work and efficiency as functions of the coupling strength for different values of $U$,
the results presented in Fig.~\ref{otto_work and eifficiency}(a) and Fig.~\ref{otto_work and eifficiency}(b) demonstrate that as $U$ increases, the peak of the output work gradually increases, whereas the corresponding coupling strength $\lambda$ gradually decreases. This behavior is advantageous for the fabrication of heat engines with high output work. In particular, when $U$ is positive, the output work of the heat engine can be optimized.
The efficiency curves in Figs.~\ref{otto_work and eifficiency}(c) and~\ref{otto_work and eifficiency}(d) indicate that positive $U$ values increase efficiency in the low coupling strength region but reduce peak efficiency. While a negative $U$ increases peak efficiency, the required coupling $\lambda$ becomes stronger.
The results in Fig.~\ref{otto_work and eifficiency} provide crucial guidance for the design of heat engines, enabling performance optimization through rational regulation of $U$ and $\lambda$.

To clarify the control mechanism of the parameter \(U\), based on the relationship among the coupling strength, efficiency, and work shown in Fig.~\ref{otto_work and eifficiency}, we identified the extrema and zero points. By using the superradiant phase transition formula 
\begin{equation}~\label{equ:lambda_c(T)}
\lambda _c(T) = \sqrt{\frac{\Delta }{4}\left[ {\frac{{\omega  }}{{\tanh(\frac{\Delta }{{2{k_B}T}})}} - \frac{{U }}{2}} \right]}.
\end{equation}
 We can accurately determine the superradiant phase transition points of the DS model, which are marked in Table \ref{table for mass and decay width}.
The chart shows that the coupling strengths corresponding to the zero points and extrema appear near the superradiant phase transition points, and as \(|U|\) increases, the degree of agreement between the zero points and phase transition points further increases.

\begin{table*}
		\begin{ruledtabular}
			\centering
			\renewcommand{\arraystretch}{1.4}
			\caption{Relationship among the hyperradiative phase transition point, efficiency, and work done. The superradiative phase transition is given by formula (1), with $T=0.1$ and $ \omega_c=\omega$. }
			\begin{tabular}{c c c c c c c c c c}\label{fig12}
			Stark intensity & $U=-0.9$ & $U=-0.6$ & $U=-0.3$ & $U=-0.1$ & $U=0$   & $U=0.1$ & $U=0.3$ & $U=0.6$ & $U=0.9$ \\
            SPT $(\lambda)$
                & 0.602  & 0.570  & 0.536  & 0.512  & 0.500 & 0.487 & 0.461 & 0.418 & 0.370 \\
			\hline
			efficiency peak$(\lambda)$     & 0.574  & 0.548  & 0.517  & 0.494  & 0.481 & 0.468 & 0.440 & 0.387 & 0.323 \\
            efficiency zero$(\lambda)$
& 0.601  & 0.577  & 0.550  & 0.529  & 0.518 & 0.507 & 0.479 & 0.432 & 0.372 \\
\hline
work peak$(\lambda)$           & 0.502  & 0.470  & 0.439  & 0.416  & 0.401 & 0.386 & 0.350 & 0.29  & 0.200 \\
work zero$(\lambda)$
                & 0.600  & 0.577  & 0.549  & 0.528  & 0.517 & 0.504 & 0.478 & 0.43  & 0.370 
			\end{tabular}
			\label{table for mass and decay width}
		\end{ruledtabular}
	\end{table*}

\begin{figure*}[t]
\includegraphics{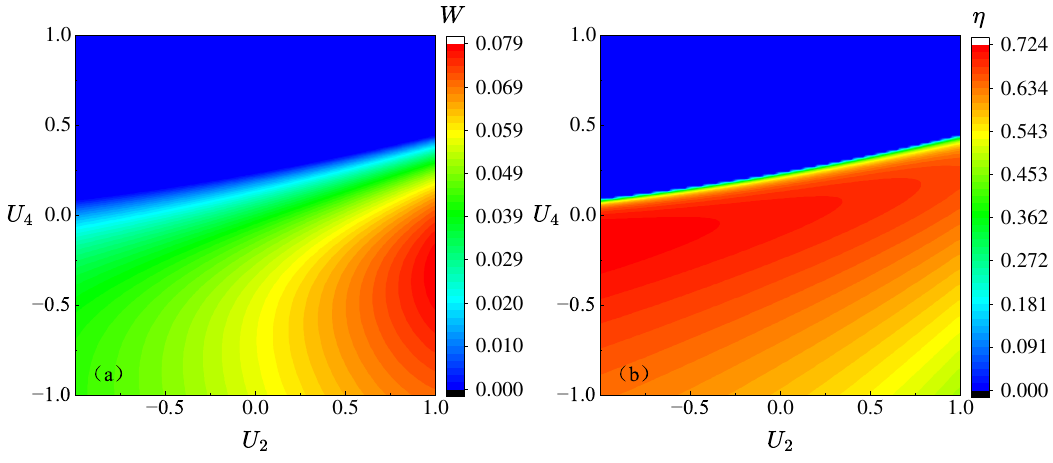}
\caption{Based on the two-dimensional heatmap analysis of work output (a) and efficiency (b) for the DS model quantum Otto cycle, the horizontal axis is set to the second stroke Stark field strength $U_2$, and the vertical axis is set to the fourth stroke Stark field strength $U_4$, with the right-side color bar quantifying the work output and efficiency values. The heatmap reveals that within the control range of the Stark interaction field strength $U_{2(4)} \in [-1, 1]$, the heat engine's performance evolves with the Stark field strength of both strokes. A significant feature is the enhancement effect on work output due to the asymmetric configuration of $U_2$ and $U_4$. The additional system parameters are $\lambda=0.48$, $N = 8$, $\omega _h = 2\omega$, $\omega _c = \omega$, $\omega  = 1$, $T_c = 0.1$, and $ T_h = 0.5$. }~\label{otto_Asymmetric Stark stroke}
\end{figure*}
\begin{figure*}[htp]
\includegraphics{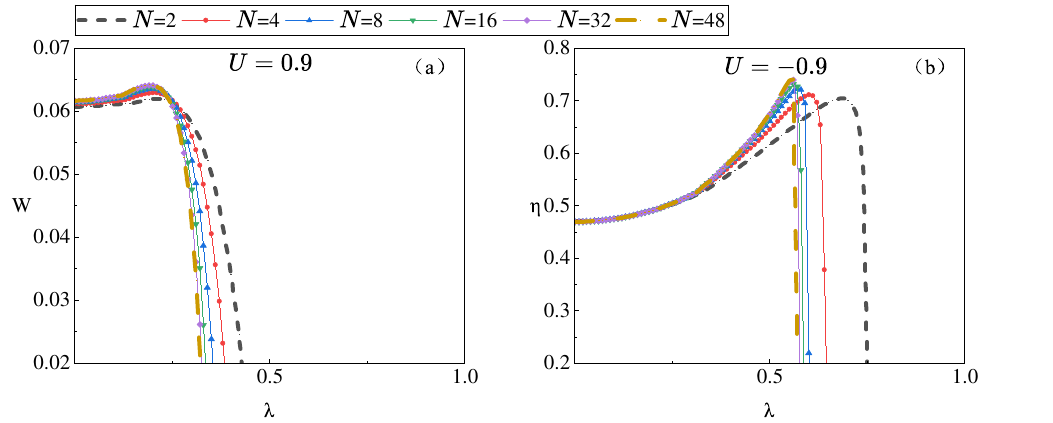}
\caption{ Figures(a) and (b) Evolution of the work output ($W$) and efficiency ($\eta$) of the quantum heat engine with coupling strength ($\lambda$) for different numbers of atoms ($N$). The horizontal axis represents the coupling strength ($\lambda$), and the vertical axis represents the work output ($W$) and efficiency ($\eta$). The multicolored curves correspond to experimental data for different atom numbers, revealing the differences in system performance under finite-size effects. In Fig. (a), U=0.9, in Fig. (b), U=-0.9, and the additional system parameters are $\omega _h = 2\omega$, $\omega _c = \omega$, $\omega  = 1$, $T_c = 0.1$, and $ T_h = 0.5$.}\label{otto_Atoms}
\end{figure*}

To further improve the performance of heat engines, we investigated an asymmetric quantum heat engine that applies distinct control parameters $U$ during isochoric expansion (second stroke) and isochoric compression (fourth stroke) \cite{shastri2022optimization}. In Fig.~\ref{otto_Asymmetric Stark stroke}, the optimal work output and efficiency do not occur in symmetric $U$ configurations (i.e., along the diagonal $U_2=U_4$ in Fig.~\ref{otto_Asymmetric Stark stroke}), highlighting the potential of asymmetric modulation $U$ to increase the power and work output. This asymmetric optimization originates from the differential sensitivity of the population distributions to $U$ across temperature regimes. As shown in Fig.~\ref{otto_Asymmetric Stark stroke} (a), the output of work peaks when a positive $U$ is applied in the second stroke and a negative $U$ is applied in the fourth stroke. The efficiency reaches its peak when a negative $U$ is applied in the second and fourth strokes, as illustrated in Fig.~\ref{otto_Asymmetric Stark stroke}(b). The divergent control requirements for maximizing work output versus efficiency demonstrate the impossibility of simultaneously optimizing both performance metrics \cite{cavina2017slow}.

In the finite-size DS model, the effect of the number of atoms ($N$) on the performance of the quantum heat engine must be investigated.
To explore the limited performance of the heat engine, we focus on the peak values of work output and efficiency. Figure.~\ref{otto_Atoms}(a) reveals that the maximum output work occurs in a system with $N=2,8,16,32,48$ and $U=0.9$. The system with $N=48$ achieves the maximum output work. As $N$ increases, the peak of the output work increases slightly, and the values of the maximum output work for $N=16,32,48$ are almost identical. Figure.~\ref{otto_Atoms}(b) reveals the maximum efficiency at $U = -0.9$ and the same $N$ value as that in Fig.~\ref{otto_Atoms}(a). The maximum efficiency increases with increasing $N$, and at $N > 32$, the maximum efficiency is almost no longer affected by the number of atoms $N$.
Although  both the maximum efficiency and maximum output work increase with the number of atoms in this figure, this occurs at a specific $U$. Different values of $U$ may lead to the maximum values being reached at different numbers of atoms. Therefore, in the above discussion, selecting $N=8$ can balance the optimization of the power output and efficiency of the heat engine.

\section{\label{sec:level5}Finite-time Cycle}

\subsection{\label{sec:level5-1}Work and Heat Calculations}
The optimization of quantum heat engines in finite-time operation has attracted significant attention in recent studies, particularly with respect to both increased efficiency and increased power output
\cite{ma2018universal,ma2018optimal,dann2020fast,barrios2021light}.
Theoretically, for the same work output, reducing the cycle duration leads to higher power. However, rapid isochoric expansion and compression strokes cause entropy generation and quantum friction, making the optimization of stroke duration a challenging issue.

To simulate a finite-time Otto cycle, we rigorously derived expressions for work and heat exchange under finite-time constraints from fundamental thermodynamic principles\cite{peterson2019experimental}.
\begin{equation}\label{eq:W(t)}
W(t) = \int_0^t d {t^\prime }{\rm Tr}\left\{ {\dot H({t^\prime })\rho ({t^\prime })} \right\},
\end{equation}\label{eq:Q(t)}
\begin{equation}
Q(t) = \int_0^t d {t^\prime }{\rm Tr}\left\{ {H({t^\prime })\dot \rho ({t^\prime })} \right\}.
\end{equation}

Corresponding to the four strokes of the heat engine in Fig.~\ref{otto_Otto cycle}, the duration of each stroke is denoted ${\tau _n} $, where $n=1,2,3,4$.
The two isochoric processes have the same evolution time ${\tau _1} = {\tau _3}$, during which the system exchanges heat with the heat reservoirs without performing any work. During the isochoric heating process, the system absorbs heat from the high-temperature reservoir, evolving from $\rho(T_c)$ to $\rho(T_h)$ within $\tau_1$. Conversely, during the isochoric cooling process, the system releases heat to the low-temperature reservoir, evolving from state $\rho(T_h)$ to $\rho(T_c)$ within $\tau_3$.
The two adiabatic processes have the same evolution time ${\tau _2} = {\tau _4}$, with no heat exchange occurring and only work being done. In the adiabatic expansion process, the frequency of the system evolves linearly from $\omega_h$ to $\omega_c$ within $\tau_2$; the reverse occurs in the adiabatic compression process.

Usually, the evolution of the system's state is regarded as a quasistatic process, where the system is in a thermal equilibrium state described by Eq.~\eqref{eq:master_equation} at any moment. However, within a finite-time framework, the quasistatic approximation is no longer applicable, and the time evolution of the system's state is governed by the master equation described by Eq.~\eqref{eq:rho}.
The total entropy generated during a complete cycle of a heat engine is often used to evaluate the impact of finite-time operations on the efficiency of the heat engine. For an Otto cycle heat engine, the total entropy can be expressed as\cite{umegaki1954conditional,vedral2002role,hayashi2017quantum}
\begin{equation}
\begin{array}{r@{}l}
\langle \sum_{tot} \rangle   =&  -\frac{1}{T_h}\left[ {{\rm Tr}({\rho _{{t_1}}}{H_h}) - {\rm Tr}({\rho _{{t_4}}}{H_h})} \right]\\
 &-\frac{1}{T_c}\left[ {{\rm Tr}({\rho _{{t_3}}}{H_c}) - {\rm Tr}({\rho _{{t_2}}}{H_c})} \right].\\
\end{array}
\end{equation}
The efficiency of the quantum engine can be related to the total entropy produced in a cycle as
\cite{xiao2023thermodynamics,peterson2019experimental,kosloff2017quantum},
\begin{equation}~\label{eq.entropy}
    {\eta _{th}} = \eta _c^{} - \frac{T_c{\langle \sum_{tot} \rangle}}{ {{\rm Tr}({\rho _{{t_1}}}{H_h}) - {\rm Tr}({\rho _{{t_4}}}{H_h})} },
\end{equation}
where ${\eta _c}=1-\frac{T_c}{T_h}$ represents the efficiency of the Carnot cycle.
\subsection{\label{sec:level5-2}Quantum Friction}
In a finite-time quantum Otto cycle, nonadiabatic transitions during compression and expansion strokes generate quantum friction, which converts useful work to dissipated heat\cite{kosloff2017quantum,abah2019shortcut}. The friction work $W_{\text{fric}}$ is defined as follows:
\begin{equation}
 W_{\text{fric}}=W_{\text{limited}}-  W_{\text{actual}},
\end{equation}
where $W_{\text{actual }}$ is the actual work output and where $W_{\text{limited}}$ can be calculated via Eq.~\eqref{eq:W(t)}.
According to the entropy-based framework \cite{fogarty2020many}, $W_{\text{fric}}$ can be quantified by the quantum relative entropy between the actual state $\rho$ and the adiabatic reference state $\rho^{\text{ref}}$:
\begin{equation}
W_{\text{fric}} ={T_{h(c)}}D\left( \rho_{2(4)} \parallel \rho_{2(4)}^{\text{ref}} \right),
\label{eq.friction}
\end{equation}
where $\quad D(\rho \parallel \rho^{\text{ref}}) = \operatorname{Tr} \left[ \rho (\ln \rho - \ln \rho^{\text{ref}}) \right]$ and where $\rho^{\text{ref}}_{2(4)}$ represents the instantaneous equilibrium state in the compression (expansion) process. Relative entropy $D$ measures information loss due to finite-time driving. Nonadiabatic transitions between instantaneous eigenstates cause an increase in $D$, directly reflecting friction-induced work dissipation.

\subsection{\label{sec:level5-3}Fidelity in Finite-time Cycles}
\begin{figure}[htb]
\includegraphics[width=0.48\textwidth]{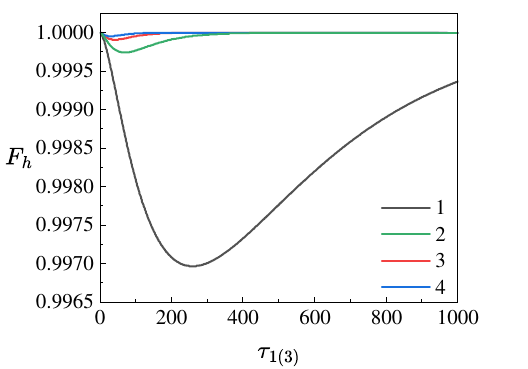}
\caption{Fidelity for
Fidelity between $\rho_{\tau_0}^{m-1}$ and $\rho_{\tau_0}^m$ as a function of the stroke time $\tau_{1(3)}$, where the number of iterative cycles $m = 1,2,3,4$ corresponds to the solid lines in black, red, blue, and green, respectively. The additional system parameters are $N = 2$, $\lambda=0.47$, $\omega _h = 2\omega$, $U=0$, $\omega _c = \omega$, $\omega  = 1$, $T_c = 0.1$, and $ T_h = 0.5$. }\label{otto_Fidelity}
\end{figure}

Under finite-time constraints, thermodynamic cycles typically operate in a nonequilibrium state, causing the heat engine's trajectory to deviate from the equilibrium state, which has detrimental effects on the operational stability of the heat engine.
Therefore, the evolution of the system state during the multiple cycles of a heat engine must be studied. The system starts the first stroke with a hot Gibbs state $\rho_{\tau_0}$ as the initial state, and after a cycle is completed, it enters the final state $\rho_{\tau_4}$ of the fourth stroke. This final state then serves as the starting point $\rho_{\tau_0}$ for subsequent cycles, and the operation process continues. After multiple iterations of the cycle are performed, the fidelity of the system state can be defined as
\begin{equation}
{F_h} ={\rm Tr}\sqrt {\sqrt {{\rho ^{(m - 1)}}({\tau _0})} {\rho ^{(m)}}({\tau _0})\sqrt {{\rho ^{(m - 1)}}({\tau _0})} } ,
\end{equation}
which measures the stability of the heat engine.
We calculated the fidelity of the system state within 4 cycles of the heat engine by assuming that the duration of the first or third stroke is finite. As shown in Fig.~\ref{otto_Fidelity}, as the number of cycles increases, the fidelity quickly approaches 1. After four cycles, when $\tau_{1(3)}> 200$, the heat engine tends to stabilize.
Therefore, all subsequent simulations were based on five complete cycles, which are sufficient to ensure stable operation of the heat engine.

\subsection{\label{sec:level5-4}Output Work, Efficiency and Power in the Finite-time Cycle}
In this study, we fixed the number of atoms $N=2$ (since the computational workload under the finite-time framework is extremely heavy, making it difficult to meet the computational requirements for $N > 2$) and focused on three key values of Stark strength: $U=0,-0.9$ and $0.9$. Here, $U=0.9$ corresponds to the maximum output work, whereas $U=-0.9$ corresponds to the highest efficiency. To explore the performance limits of the thermal engine under finite-time operation, we employed the optimized coupling strength, $\lambda  = 0.47$ for $U = 0$, $\lambda = 0.21$ for $U = 0.9$, and $\lambda  = 0.68$ for $U = - 0.9$, which maximizes the output work and efficiency under quasistatic conditions.
\begin{figure*}[htp]
\includegraphics{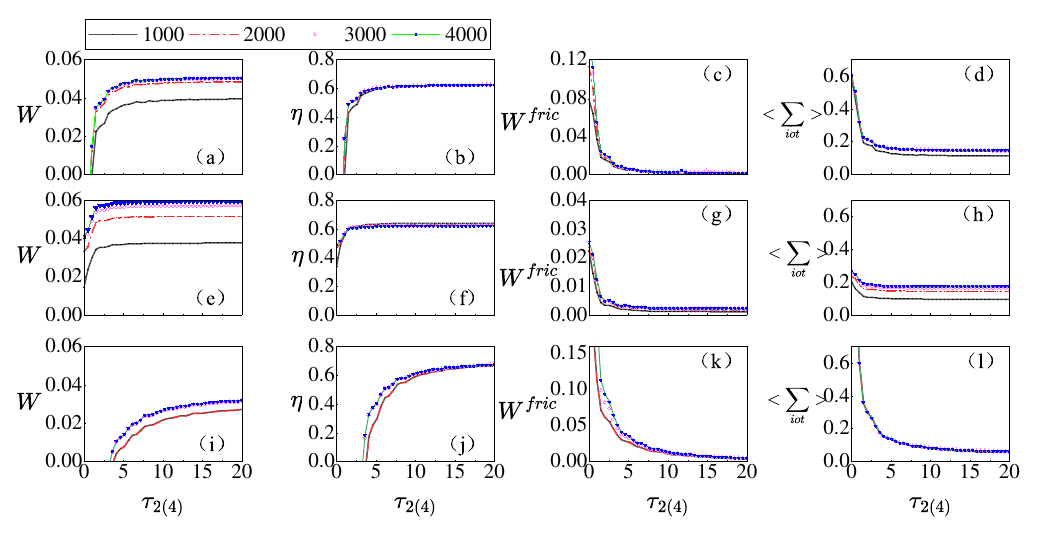}
\caption{Parameters that characterize the performance of the heat engine: work output (first column), engine efficiency (second column), quantum friction (third column), and entropy production (fourth column) as functions of the adiabatic driving time $\tau_2 = \tau_4$, whereas curves with different colors correspond to different durations of the isochoric thermalization time $\tau_1 = \tau_3$. Parameters used in the calculations: first row (a)-(d) $\lambda  = 0.47$, $U = 0$; second row combination (e)-(h) $\lambda = 0.21$, $U = 0.9$; and third row (i)-(l) $\lambda  = 0.68$, $U = - 0.9$. Other parameters include $\omega _h = 2\omega$, $\omega _c = \omega$, $\omega  = 1$, $T_c = 0.1$, and $T_h = 0.5$.}~\label{otto_Limited time work}
\end{figure*}
Under finite-time conditions, incomplete thermalization during the isochoric process and nonadiabatic excitation transitions during adiabatic strokes lead to entropy generation and quantum friction. To quantify these irreversible effects, entropy production (Eq.~\eqref{eq.entropy}) and quantum friction (Eq.~\eqref{eq.friction}) are used as performance metrics for the heat engine. As shown in
Fig.~\ref{otto_Limited time work}(c), (d), (g), (h), (k), and (l), both quantum friction and entropy production monotonically decrease with increasing stroke duration $\tau_{2(4)}$ and gradually stabilize after their minimum values are reached. Figures. \ref{otto_Limited time work}(a), (b), (e), (f), (i), and (j) indicate that the output work of the system increases monotonically with increasing stroke duration $\tau_{2(4)}$ and gradually reaches saturation.
These results demonstrate that by optimizing the Stark strength $U$ and interaction strength $\lambda$,
when the isochoric process time $\tau_{1(3)}$ is between 1000 and 4000 and the adiabatic process time $\tau_{2(4)}<20$, the heat engine achieves stable performance after five working cycles are completed.
Comparing each column of the plots for $U = 0,0.9$ and $-0.9$ in Fig.~\ref{otto_Limited time work} reveals that as $U$ increases, the heat engine requires a longer adiabatic process time $\tau_{2(4)}$ to achieve a stable operating state and effectively reduces quantum friction.

In particular, Figs.~\ref{otto_Limited time work}(a), (e), and (i) show that when the isochoric process time $\tau_{1(3)}$ is set to 1000, 2000, 3000, or 4000, the corresponding output work $W$ of the heat engine increases.
\begin{figure*}[htp]
\includegraphics{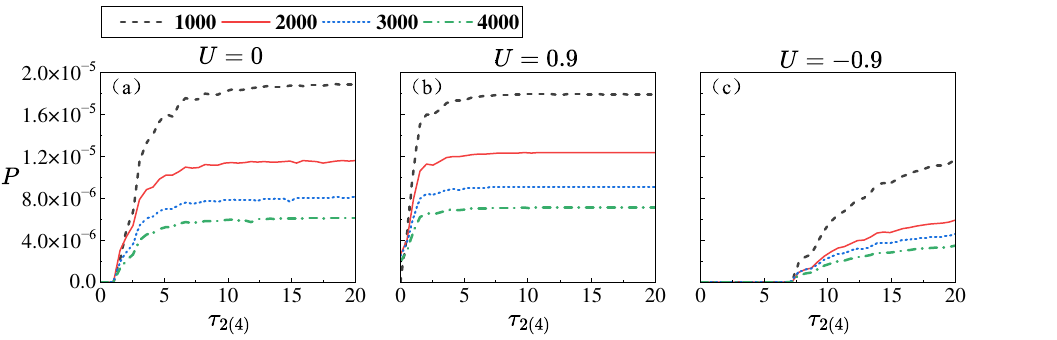}
\caption{Power curves as a function of the adiabatic stroke time $\tau_{2(4)}$, with the isochoric stroke times $\tau_{1(3)}=1000, 2000, 3000$ and $4000$ corresponding to the black shot, red, blue and green lines, respectively. The Stark strength $U$ and coupling strength $\lambda$ in (a), (b) and (c) are $\lambda  = 0.58$ for $U = 0$, $\lambda  = 0.21$ for $U = 0.9$ and $\lambda  = 0.68$ for $U =  - 0.9$, respectively. Other parameters include $\omega _h = 2\omega$, $\omega _c = \omega$, $\omega  = 1$, $ T_c = 0.1$, and $T_h = 0.5$. }\label{otto_Power}
\end{figure*}
Under finite-time conditions, power is a key parameter characterizing the performance of Otto heat engines and is influenced by the adiabatic stroke time $\tau_{2(4)}$, isochoric stroke time $\tau_{1(3)}$, Stark field strength and coupling strength $\lambda$, as illustrated in Fig.~\ref{otto_Power}. As $\tau_{2(4)}$ increases, the output power gradually increases and reaches saturation. Moreover, when $\tau_{1(3)} = 1000, 2000, 3000$ and $4000$, the values at which the power saturates decrease successively, which is contrary to the results of the output work shown in Fig.~\ref{otto_Limited time work}.
When $(U = 0$ and $U = 0.9$, the power of the heat engine is significantly greater than that when $U = -0.9$. The maximum power output occurs at $U = 0$ and $\lambda = 0.58$ [see the black solid line in Fig.~\ref{otto_Power}(a)]. For $U = 0.9$ and $\lambda = 0.21$, the output power shown in Fig.~\ref{otto_Power}(b) is nearly identical to that shown in Fig.~\ref{otto_Power}(a). Crucially, modulating the Stark intensity achieves similar output power, while the coupling strength $\lambda$ is significantly reduced, which is beneficial to the design of heat engines.

\begin{figure*}[t]
\includegraphics{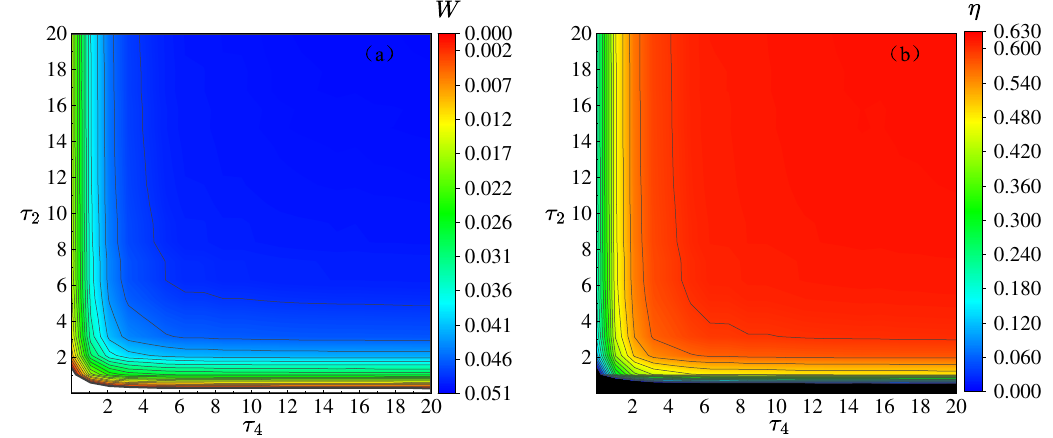}
\caption{ (a) and (b) Two-dimensional heatmap distributions of the output work (a) and efficiency (b) of the asymmetric DS quantum Otto heat engine in the $\tau_{2} -\tau_{4} $ stroke time space. The horizontal axis represents the duration of the fourth stroke \(\tau_{4} \), whereas the vertical axis represents the duration of the second stroke \(\tau_{2} \); the color scale quantifies the performance. The additional system parameters are $\lambda=0.48$, $U=0$, ${\omega _h} = 2\omega$, ${\omega _c} = \omega$, $\omega  = 1$, $\ T_c = 0.1$, and $\ T_h = 0.5$.}~\label{otto_Asymmetric time stroke}
\end{figure*}

There exists a time-asymmetric stroke heat engine model \cite{zheng2016occurrence, shastri2022optimization} characterized by independent adjustment of the durations of the two adiabatic strokes. Using the duration of the second adiabatic stroke $\tau_{4}$ as the horizontal axis and the duration of the fourth adiabatic stroke $\tau_{2}$ as the vertical axis, the output work and efficiency exhibit obvious asymmetry.
As shown in Fig.~\ref{otto_Asymmetric time stroke}(a), the peak output work deviates markedly from the diagonal, exhibiting a stronger dependence on the duration of the second adiabatic stroke $\tau_2$, and appears in the region where $\tau_2$ is relatively small.
The efficiency peak also displays asymmetry and depends more on the duration of the fourth adiabatic stroke $\tau_4$, as shown in Fig.~\ref{otto_Asymmetric time stroke}(b). Consequently, targeted adjustment of the time allocation between these adiabatic strokes maximizes both the power output and the thermodynamic efficiency.

\section{\label{sec:level6}Conclusion}
Numerical simulations of a quantum Otto engine were performed using the DS model as the working substance. The effects of the Stark field strength, coupling strength, adiabatic stroke duration, isochoric stroke duration, and number of atoms in the DS model on the heat engine's output work, efficiency, and power output were investigated.
Within the infinite-time thermodynamics framework, regulating the Stark field strength and coupling strength enables conversion between four distinct operating modes and allows optimization of the heat engine's operating state to maximize output work and efficiency. The calculation results show that the maximum values of output work and efficiency occur near the coupling strength $\lambda_c$ corresponding to the superradiant phase transition. Within the range of [-0.9, 0.9], a larger $U$ can increase the heat engine's output work, whereas a smaller $U$ is beneficial for enhancing its efficiency. Applying different Stark field strengths $U_2$ and $U_4$ to control the adiabatic expansion and compression processes facilitates optimization of the heat engine performance. Additionally, in the DS model, an increase in the number of atoms also helps to increase the output work and efficiency of the heat engine. A complete cycle of an actual heat engine takes a finite amount of time. Within the infinite-time thermodynamics framework, the output work and efficiency of the heat engine are less than those under infinite-time conditions, which is due to entropy generation and quantum friction arising during the nonequilibrium evolution of the system's quantum state. By adjusting the isochoric stroke time $\tau_{1(3)}$, adiabatic stroke time $\tau_{2(4)}$, Stark field strength and coupling strength, the heat engine can achieve stable output work, efficiency, and power after 5 complete cycles. The increases in $\tau_{1(3)}$ and $\tau_{2(4)}$ cause the output work and efficiency to reach saturation. A reduction in $\tau_{1(3)}$ is beneficial for increasing power. A larger $U$ makes it possible to design a heat engine with relatively high efficiency under conditions of low coupling strength. Moreover, a time-asymmetric heat engine with independent regulation of the adiabatic compression time $\tau_2$ and adiabatic expansion time $\tau_4$ is favorable for optimizing the output work and efficiency of the heat engine.
The work presented in this paper contributes to the design of high-performance quantum heat engines.

\section*{Acknowledgments}

We acknowledge the useful discussions with Jiasen Jin and He-Guang Xu. This work is supported by the Science and Technology Projects of the China Southern Power Grid (YNKJXM20220050).

\bibliographystyle{unsrt}
\bibliography{APS_DSOTTO}
\end{document}